# Visual Environment for Rapid Composition of Parameter-Sweep Applications for Distributed Processing on Global Grids

S. Burq[1], S. Melnikoff[1], K. Branson[2], and R. Buyya[1,*]

[1] Grid Computing and Distributed Systems Laboratory
Dept. of Computer Science and Software Engineering
The University of Melbourne, Australia

[2] Structural Biology
Walter and Eliza Hall Institute (WEHI)
Parkville, Melbourne, Australia

**Abstract**

Computational Grids are emerging as a platform for next-generation parallel and distributed computing. Large-scale parametric studies and parameter sweep applications find a natural place in the Grid's distribution model. There is little or no communication between jobs. The task of parallelizing and distributing existing applications is conceptually trivial. These properties of parametric studies make it an ideal place to start developing integrated development environments (IDEs) for rapidly Grid-enabling applications. However, the availability of IDEs for scientists to Grid-enable their applications, without the need of developing them as parallel applications explicitly, is still lacking. This paper presents a Java based IDE called Visual Parametric Tool (VPT), developed as part of the Gridbus project, for rapid creation of parameter sweep (data parallel/SPMD) applications. It supports automatic creation of parameter script and parameterisation of the input data files, which is compatible with the Nimrod-G parameter specification language. The usefulness of VPT is demonstrated by a case study on composition of molecular docking application as a parameter sweep application. Such applications can be deployed on clusters using the Nimrod/enFuzion system and on global Grids using the Nimrod-G grid resource broker.

## 1. Introduction

As high-speed networks become ubiquitous and research in middleware technologies matures, new windows of opportunity for application scientists to run their applications on parallel and distributed computing environments, such as clusters and Grids [3], are increasing. The underlying infrastructure, providing the low-level facilities to run applications in a heterogeneous and distributed environment, makes up part of the Grid. The Grid also includes higher-level tools and applications for the various interest groups involved in the development.

There exist a number of models for the construction of parallel and distributed applications. Parameter sweep is one of the simplest and most practical of the models that can yield powerful results. From large matrix-multiplication to drug discovery via molecular docking, parametric studies find a use in most sciences. Some application scenarios include:
- geologist looking at the change in the density and depth of ore-body and the overlying rock's density to optimize cost and production;
- molecular biologist looking for a chemical, in a large set, that best fits a particular protein.
- aerospace engineer understanding the role of geometry parameters in the aerodynamic design and optimization process.

The practical implications of performing parametric studies make it difficult for an application scientist, who has little or no knowledge of distributed computing, to use it effectively. The vision of the Grid is precisely to bridge this gap by providing scientists, all over the world, seamless access to compute and other scientific resources without them worrying about the lower-level details of the computing

---

infrastructure or the resource management and discovery issues involved in doing so [1]. Tools that will provide an interface to the Grid make up an essential part of this vision. These must be easy to understand and use. Currently, the availability of integrated development environments (IDEs) with visual interface for scientists to rapidly Grid-enable their existing applications is still lacking.

This paper presents a Java based IDE called Visual Parametric Tool (VPT), developed as part of the Gridbus project, for rapid creation of parameter sweep (data parallel/SPMD) applications. VPT provides a simple interface for the manipulation of scripts or input files of existing applications. Users can visually assign parameters to certain values by highlighting them. They can select from a number of different data types and domains to describe their parameters. VPT also consists of a task editor. This is for creating the tasks performed during different stages of a distributed execution. The parameters and tasks together provide the basis of each run. VPT allows the rapid creation and manipulation of the parameters. While being flexible, it is also simple enough for a non-expert to create a parameter script, known as a plan file, within minutes. The parameter sweep applications composed using VPT can be deployed on global Grids using the Nimrod-G resource broker that supports scheduling based on the user's quality of service (QoS) requirements—such as the deadline, budget, and optimization preference—and the access price of resources.

The rest of this paper is organised as follows. Section 2 presents related tools and their capabilities including differences. We will then examine architecture in Section 3 and the design and implementation details of VPT in Section 4. The use of VPT for composing molecular docking application as a parameter sweep application is discussed in Section 5, followed by a conclusion in Section 6.

## 2. Related Work

VPT draws inspiration from or builds on the concepts developed in Nimod [2] and its commercial version (Enfuzion [1]); and its Grid-enabled version (Nimrod-G [9]) that support the creation and execution of parametric applications on clusters and Grids respectively. A declarative language, called parameter specification language, supported by Nimrod describes the parameters and the tasks that make up the plans.

For the creation of plans, Enfuzion takes a wizard approach. Enfuzion will take a user through the operation of creating a job specification file step-by-step, because the operation is a bit complex for novice users to create parameter script on their own. This keeps the parameterization process linear and prevents errors. In the input file to the application, the user must change the value assigned to a parameter to a place marker. Although simple and less prone to error, this approach is too rigid, slow and cumbersome for someone working on several input files at the same time. As the parameter script and parameterized input data files generated by VPT confirm to the Nimrod parameter specification language, it serves as a complimentary tool. This ensures that VPT can be leveraged by Nimrod, EnFuzion, and Nimrod-G users.

We have taken a more application-oriented approach. The user can open a saved or a new VPT project. To this, the user can add input files. There are essentially two methods for creating new parameters. The user can, drag and select the value in the input file that they wish to assign a parameter to, or they can create parameters independent of an input file. This gives the user a great deal of flexibility and control. By giving the user fields to input their parameter configuration and then generating the plan specification automatically we can prevent errors. The user also has access to normal editing tools. Even if they create parameter script in their favorite editor, VPT allows them to import and make use of its capabilities. This approach has a number of advantages:
- It is faster because the user only changes those things that the program got wrong.
- The users do not have to bother with attributes and features they are not interested in it.
- The user can save a project and modify it at a later stage.

Other related works include, APST (AppLeS Parameter Sweep Template) [10] and IPG's parameter process specification tool [11]. APST expects application scientists to explicitly create jobs, which can then be deployed on the Grid by its scheduler. Although IPG provides graphical environment for parameterising the data files, but its integrated scheduler is uses traditional system centric policy in resource allocation. As VPT confirms to the Nimrod-G parameter specification language, it enables the users to harness Grid resources using the Nimrod-G resource broker depending on their QoS requirements and the access price of resources. Thus, it supports the Grid economy, which is essential for management and allocation of resources based on the supply and demand.

## 3. Architecture

VPT consists of three major visual components: Project, Input Files and Tasks. These components are represented as Project Window, Input File Window and Task Editor. The design of VPT, shown in Figure 1, allows a single project to have several input data files and tasks.

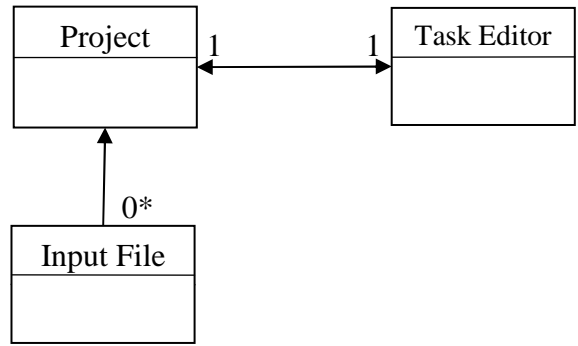

**Figure 1:** Basic visual components of VPT.

These visual components provide the user access to the objects that encapsulate the plan's information-model, namely to ParamObject and GBTask. ParamObject is created and manipulated from the project window or input file window, while the GBTask is created and manipulated using the TaskEditor.

A plan consists of parameters and task. In VPT, parameters are internally represented as ParamObjects and tasks as GBTasks. ParamObjects are created by any of the following three methods.
- File dependent parameterization
- File independent parameterization
- Via imported plan specification

*File dependent parameterization*
Once an input file or a script file is imported into VPT, values that have to be assigned parameters are highlighted and the parameter defined and assigned by a simple click of mouse (see Figure 2).

*File independent parameterization*
New parameters may also be created by simply defining its properties (see Figure 2).

*Via imported plan specification*
VPT contains a LALR parser for plan specification based on the grammar in Figure 3. This allows the reuse of an existing plan file (parameter script). The parser translates each parameter definition into a ParamObject and each task description into a GBTask (see Figure 2).

*Experiment editor and job generation*
Once a plan specification is completed, VPT can generate a run specification. This states every value lying within the range of the parameters described by the plan, and a description of the jobs in terms of the values assigned to them. Hence, the run specification describes the distribution model of the application parameterized using VPT.

## 4. Design and Implementation

VPT is coded in java. Besides the above-mentioned objects, VPT has various components that facilitate the creation of a plan specification (parameter script) and parameterisation on input data files. The graphical user interface is written in swing. VPT is divided into following of packages.

*ExperimentEditor*
This contains the GUI classes for the ExperimentEdior. It also contains a controller class (following the classic MVC architecture) that processes the user input.

*GBJobs*
This contains a single class, `Jobs`. "Jobs" takes as its input a count (N) of those parameters that have a range of values and an array of integers of size N containing the maximum value taken by each of these parameters.

*GBTask*
This package also contains a single class, `GBTask`. It is a serializable object. It encapsulates the commands that execute during different phases of the distributed run.
`GBTask` – constructor summary:
`public GBTask(java.lang.String name)`

*GridBus*
This is the largest package containing mostly the GUI classes for VPT. Following the MVC architecture, it contains all the "view" components. It also includes a utility class, called `GBFileManager`, for handling all file operations within VPT. In addition, this package contains the class that has VPT's main method, named `Project`.

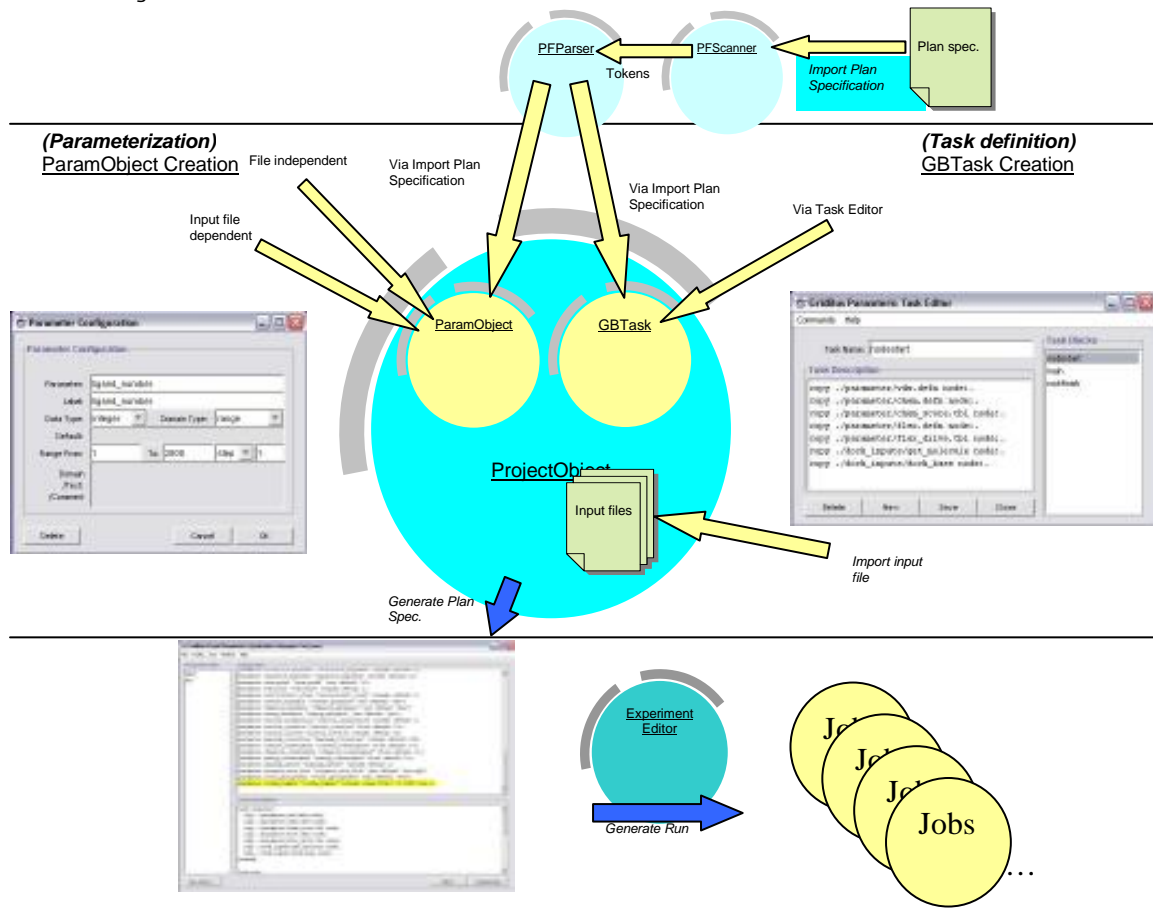

**Figure 2**: Interaction between VPT components to create plan and generate a run.

*ObserverPattern*
This package contains two interfaces *Observer* and *Subject*. This facilitates the implementation of MVC architecture, by decoupling related objects [4]. A subject may have a number of observers. All observers are notified when the subject undergoes a state change. In response, the observer may query the subject to synchronize its state with the subject. The observer implements the `update()` method while the subject implements the `addObserver()` and `removeObserver()` method. On a state change, the subject calls each observer's update method.

*ParamObject*
This package contains a single class: `ParamObject`. The `ParamObject` is the heart of VPT. It is a serializable object encapsulating the state of a parameter, it contains two key methods: `makePlanStep()` and `makeRunStep()`. These methods are responsible for automating the process of plan and run specification creation. The `makePlanStep` method converts the fields of the ParamObject into a line of the simple declarative language following the grammar of Figure 3. `makeRunStep` converts a parameter's declaration into a statement of a run specification. This declaration identifies the possible value(s) taken by the parameter. Currently `makeRunStep` generates a Nimrod-G readable statement.

$$plan \rightarrow plan\ rest\ |\ error\ |\ \varepsilon$$
$$rest \rightarrow planStep\ |\ taskBlock\ |\ newline$$
$$planStep \rightarrow PARAMETER\ ID\ label\ type\ domain\ SEMI$$
$$label \rightarrow LABEL\ QUOTE\ |\ QUOTE\ |\ \varepsilon$$
$$type \rightarrow INTEGER\ |\ FLOAT\ |\ TEXT\ |\ FILE$$
$$domain \rightarrow DEFAULT\ value\_opt$$
$$|\ RANGE\ range\_values\ domain2$$
$$|\ SELECTANY\ value\_list\ default\_opts$$
$$|\ SELECTONE\ value\_list\ default\_opt$$
$$|\ RANDOM\ range\_values\ points\_opt$$
$$|\ COMPUTE\ expr$$
$$|\ JITP\ jitp\_expr$$
$$points\_opt \rightarrow POINTS\ value\_opt\ |\ \varepsilon$$
$$default\_opts \rightarrow value\_opt\ value\_list\ |\ value\_opt$$
$$domain2 \rightarrow POINTS\ value\_opt\ |\ STEP\ value\_opt\ |\ \varepsilon$$
$$value\_opt \rightarrow ID\ |\ QUOTE\ |\ NUM$$
$$expr \rightarrow expr\ PLUS\ term\ |\ expr\ MINUS\ term\ |\ term$$
$$term \rightarrow term\ TIMES\ factor\ |\ factor$$
$$factor \rightarrow NUMBER\ |\ LPAREN\ expr\ RPAREN$$

**Figure 3:** Context free grammar for plan specification.

*PFScanner*
PFScanner, (plan file scanner) created using an open source tool called JLex [5], performs lexical analysis of the plan specification. It comes into play when the user wishes to import an existing plan specification into VPT. It interfaces with the PFParser (discussed below) providing it with a stream of identified tokens.

*PFParser*
PFParser, (plan file parser) written using an open source tool called CUP [5], interfaces with the PFScanner and attempts to match the stream of tokens to a complete parameter or task definition as described by the plan grammar in Figure 3. All caps denote the terminals. In doing so, it generates new `ParamObjects` or `GBTasks`. It contains two public methods for the retrieval of ParamObject and GBTasks: `getParams()` and `getTasks()`.

*ProjectObject*
`ProjectObject` encapsulates all the attributes necessary to describe a VPT project. It contains the `ParamObjects`, `GBTasks`, paths to input files and other attributes that uniquely identify a project.

Associations and reverse-associations between packages are shown in Figure 4. The arrow heads point at the dependent packages. Notice, a single class, Jobs, in GBJobs package, is responsible for the production of Grid enabled jobs. This can be extended to support several middleware jobs.

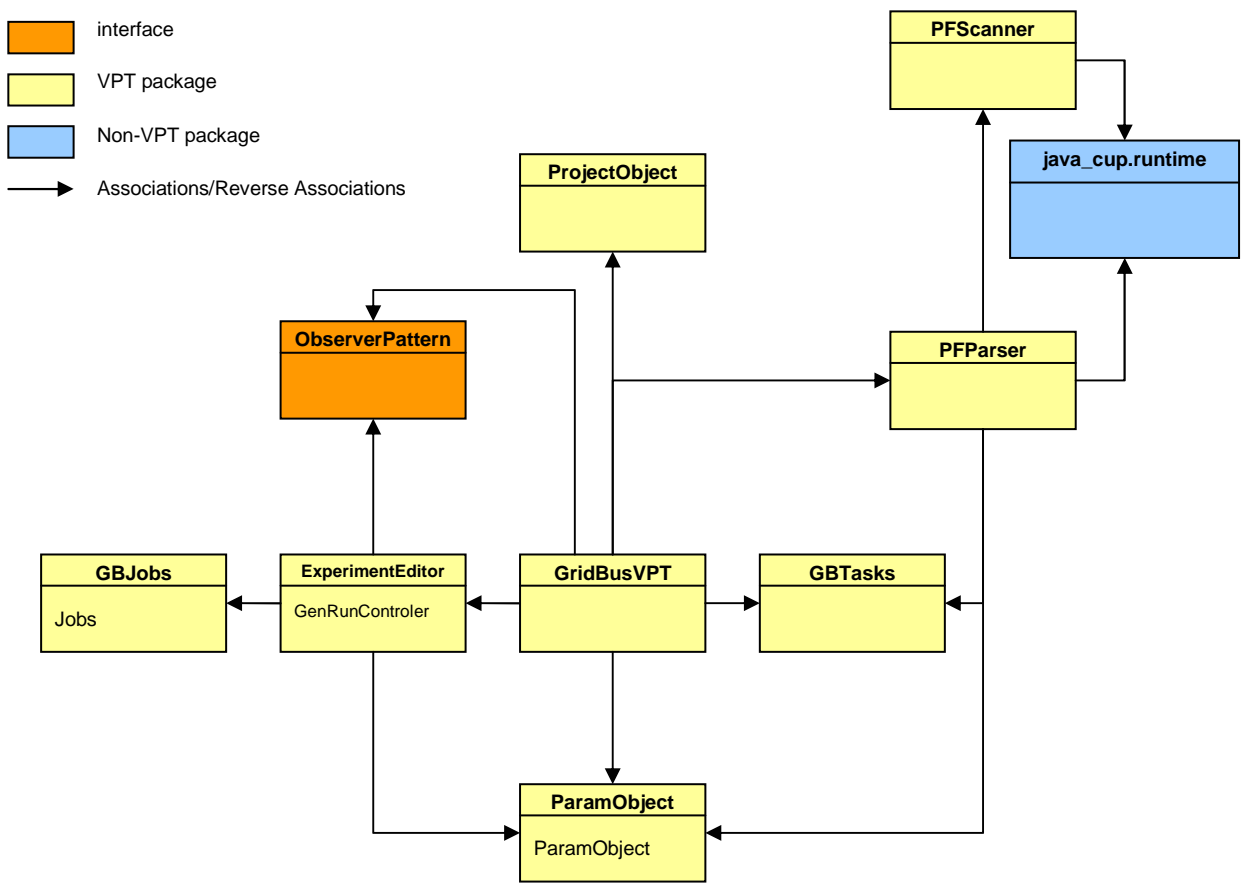

**Figure 4:** VPT package associations and reverse-associations.

## 5. Use Case Study – Molecular Docking Application

Molecular modeling for drug design involves screening millions of ligand records or molecules of compounds in a chemical database (CDB) to identify those that are potential drugs. This process is called molecular *docking* [7]. It helps scientists explore how two molecules, such as a drug and an enzyme or protein receptor, fit together. Docking each molecule in the target chemical database is both a compute and data intensive task. In [8], a virtual laboratory environment has been developed and demonstrated distributed execution of molecular docking application on Global Grids. The application has been formulated as a parameter sweep application using a simple parameter specification language and deployed on global Grids using the Nimrod-G resource broker.

We now discuss how the application has been parameterized (i.e., the creation of parameter script and parameterisation of data files) using the VPT. In [8], the creation of parameter script and parameterisation of data/configuration files has been carried out using text editors such as "vi". Although this task is simple, but it becomes cumbersome when an application contains multiple data files and has a large number of data entries to be parameterised. It also increases the likely hood of creating parameter script with syntax errors. The use of visual modeler helps in overcoming these limitations and also aids in the rapid parameterisation of the molecular docking application such as the "Dock" [7] software package.

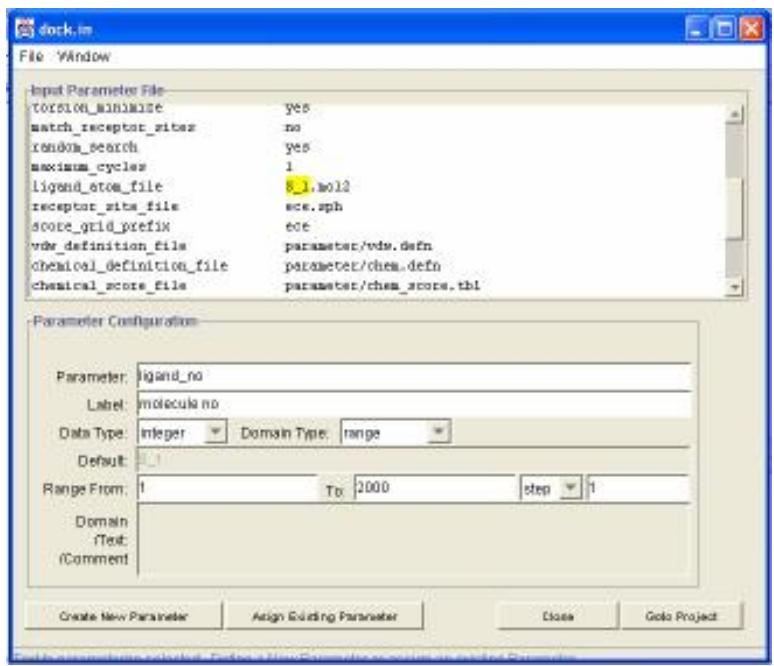

**Figure 5:** The parameterisation of docking configuration input file.

Figure 5 shows the parameterisation of docking application configuration input file using VPT. In this example, the users need to open/import this configuration input file, select the data item to be paramterized (see the highlighted text in Figure 5). Then the dialogue box appears where they can assign attributes for it. In this example, the name of a data item called "ligand_atom_file", which indicates the molecule to be screened. The attributes this parameter can be assigned as "range" type with initial values that indicate the start, end, the step value. That is, to screen the first 2000 molecules in the chemical data base, the initial values to be assigned are 1, 2000, and 1. VPT will automatically create a parameter statement and add to the script (see the highlighted statement in Figure 6). A task specification creation module provides dialogue facility selection of appropriate commands associated with the execution of a parametric job (see a small window in Figure 6).

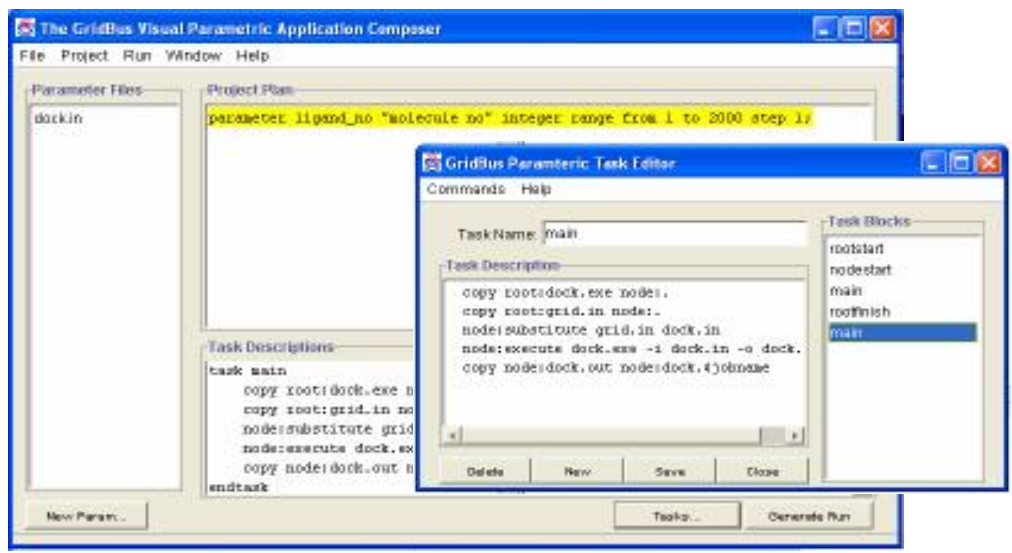

**Figure 6:** The creation of docking parameter script.

# 6. Summary and Conclusion

This paper outlined the need for the development of IDEs (integrated development environments) and other applications and tools in order to provide the applications scientist with user-friendly environments to run their code on the Grid. We introduced VPT (Visual Parametric Tool) developed to provide one such environment for parameter sweep applications. We identified its key features, while giving some of its implementation details. Thus, we showed that VPT's modular design made it possible to extend it to support other middleware technologies. Finally, we showed how application scientists can use VPT to parameterise their applications. Such paramterised applications can be deployed on Global Grids using the Nimrod-G resource broker.

## References

We thank Srikumar Venugopal and Elan Kovan for their comments on the paper.